\documentclass[onecolumn,preprintnumbers,amsmath,amssymb,superscriptaddress]{revtex4-2}
\usepackage{epsf}
\usepackage{graphicx}
\usepackage{sidecap}
\usepackage{array}
\usepackage{amsmath}
\usepackage[utf8]{inputenc}
\usepackage{amssymb}
\usepackage{dcolumn}
\usepackage{epstopdf}
\usepackage{bm}

\usepackage{color}

\def \beq {\begin{equation}}
\def \eeq {\end{equation}}
\begin{document}
 
\title{{Observation of anisotropic Dirac cones in the {topological material  Ti$_{2}$Te$_{2}$P }   }}
\author{Gyanendra~Dhakal}\affiliation {Department of Physics, University of Central Florida, Orlando, Florida 32816, USA}
 \author{Firoza~Kabir}\affiliation {Department of Physics, University of Central Florida, Orlando, Florida 32816, USA}
\author{Ashis~K.~Nandy}\affiliation {Department of Physics and Astronomy, Uppsala University, P. O. Box 516, S-75120 Uppsala, Sweden}
\affiliation{{School of Physical Sciences, National Institute of Science Education and Research,
An OCC of Homi Bhabha National Institute, Jatni-752050, India}}

\author{Alex Aperis} \affiliation {Department of Physics and Astronomy, Uppsala University, P. O. Box 516, S-75120 Uppsala, Sweden}

\author{Anup Pradhan Sakhya} \affiliation {Department of Physics, University of Central Florida, Orlando, Florida 32816, USA}
\author{Subhadip Pradhan} \affiliation{{School of Physical Sciences, National Institute of Science Education and Research,
An OCC of Homi Bhabha National Institute, Jatni-752050, India}}

\author{Klauss~Dimitri}\affiliation {Department of Physics, University of Central Florida, Orlando, Florida 32816, USA}
   \author{Christopher~Sims}\affiliation {Department of Physics, University of Central Florida, Orlando, Florida 32816, USA}
 \author{Sabin~Regmi} \affiliation {Department of Physics, University of Central Florida, Orlando, Florida 32816, USA}
 \author{M.~Mofazzel~Hosen}\affiliation {Department of Physics, University of Central Florida, Orlando, Florida 32816, USA}
 \author{Yangyang Liu}\affiliation {Department of Physics, University of Central Florida, Orlando, Florida 32816, USA}
 \author{Luis Persaud} \affiliation {Department of Physics, University of Central Florida, Orlando, Florida 32816, USA}
  \author{Dariusz~Kaczorowski}\affiliation {Institute of Low Temperature and Structure Research, Polish Academy of Sciences, Okolna 2, 5--420 Wroclaw, Poland}
\author{Peter M. Oppeneer}  \affiliation {Department of Physics and Astronomy, Uppsala University, P. O. Box 516, S-75120 Uppsala, Sweden} 
 \author{Madhab~Neupane$^*$}
\affiliation {Department of Physics, University of Central Florida, Orlando, Florida 32816, USA}
 
\date{27 October, 2017}
\pacs{}

\begin{abstract}
\noindent
  Anisotropic bulk Dirac (or Weyl) cones in three dimensional systems have  recently gained intense research interest  as they are examples of materials  with tilted Dirac (or Weyl) cones indicating the violation of Lorentz invariance. In contrast, the studies on anisotropic surface Dirac cones in topological materials which contribute to anisotropic carrier mobility  have been limited. By employing  angle-resolved photoemission spectroscopy  and first-principles calculations, we reveal the anisotropic surface Dirac dispersion in a tetradymite material Ti$_{2}$Te$_{2}$P on the (001) plane of  the Brillioun zone.  We observe the quasi-elliptical Fermi pockets at the $\bar{M}$-point of the Brillouin zone forming the anisotropic surface Dirac cones. Our calculations of the $\mathbb{Z}_{2}$ indices confirm that the system is topologically non-trivial with multiple topological phases in the same material. In addition, the observed nodal-line like feature formed by bulk bands makes this system {topologically rich}.

\end{abstract}

\date{\today}
\maketitle
The experimental discovery of a three-dimensional (3D) topological insulator (TI) in tetradymite Bi$_{2}$Se$_{3}$  brought about an unprecedented surge of  research  interests in exotic states of matter \cite{Hasan, SCZhang, Neupane3, Fu, Xia, Chen, Ando, bansil}. This discovery motivated investigations  of other novel phases such as the Dirac semimetal, Weyl semimetal, nodalline semimetal, topological crystalline insulator, Kondo insulator, etc \cite{zwang, MN_cd, BJyang, SMyoung, Weyl_science, Hweng, ZrSiS, ZrSiX, LFu, Natcommun}. Not only have the discoveries of these novel states provided new classifications of materials, but a series of compounds with similar electronic behaviors have also been identified. Recently, a great deal of interest has been directed at a series of bismuth- and antimony-based tetradymite crystals as they host a topological surface state while simultaneously possessing a wide bulk  band gap, as such a single 2D Dirac cone in the Brillouin zone (BZ) is present \cite{Xia, Chen, Ando221, CPauly, CavaMN, CKLee, Moore, MNTI, MN1, Xu1, Haijun}. Similarly, angle-resolved photoemission spectroscopy (ARPES) studies of the 221-type compounds isostructural to the tetradymite family such as Zr$_{2}$Te$_{2}$P, and Hf$_{2}$Te$_{2}$P \cite{Cava221, Niraj, Hosen221} reveal the presence of topological surface states along with multiple fermionic-states in the BZ.  In Zr$_{2}$Te$_{2}$P, multiple Dirac cones are reported, among which two Dirac cones reside within the pseudo gap. Additionally, it contains a surface state anisotropic  Dirac cone at the $\bar{M}$ point \cite{Cava221}. Another compound of this family, Hf$_{2}$Te$_{2}$P, possesses multiple {topological} states, hosting strong TI and weak TI states in a single material. Furthermore, it consists of a   one-dimensional  Dirac-node arc  along a high symmetry direction which makes this family even more fascinating \cite{Hosen221}. Interestingly, the band inversion takes place between the \textit{d} and \textit{p} bands in these compounds in contrast to other tetradymite families  which host band inversion between \textit{s} and \textit{p} bands {or between \textit{p} bands} \cite{Haijun, Cava221, Niraj, Hosen221, Niraj1}.\\
Most of the well studied TIs and topological semimetals (TSMs) have  isotropic Dirac cones which are dubbed as type-I fermions. Bi$_2$Se$_3$, Cd$_3$As$_2$ and TaAs are prime examples of {materials hosting} type-I fermions in TI, Dirac and Weyl semimetallic states, respectively \cite{Xia, MN_cd, Weyl_science}. However, very recently topological states with tilted Dirac cones have gained significant attention. Subsequently, type-II Weyl semimetal \cite{weyl1} and  type-II Dirac semimetals \cite{diracii, diracii_1, diracii_2} have been discovered in three dimensional systems. These anisotropic Dirac cones are formed by bulk bands. Type-II fermions have tilted Dirac cones as a result of broken Lorentz invariance,  and Dirac or Weyl nodes exist at the contact of electron and hole energy pockets \cite{weyl1}. In a similar fashion,  anisotropic Dirac cones have been reported in two dimensional systems in which Fermi velocities vary along high symmetry directions \cite{Feng, Hong, Zhao}.  The anisotropy of the Dirac cone might provide  anisotropic carrier mobility, thus leading to the realization of   direction-dependent transport for quantum devices \cite{Feng, Zhao}.  A few materials harboring  anisotropic surface Dirac cones have been reported  \cite{Park, Feng1, Ru2Sn3, Cava221,Virot, Zhang}. However, the  examples of 2D anisotropic Dirac cones are limited in 3D materials. \\
\begin{figure}[ht]
	\centering
	\includegraphics[width=0.95\textwidth]{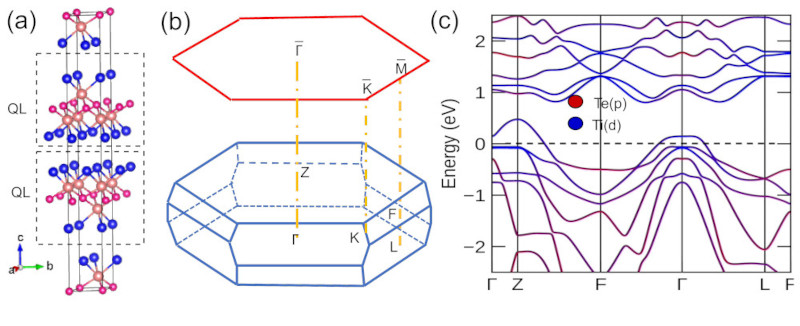}
	\caption{{Crystal structure and band structure calculations of  Ti$_{2}$Te$_{2}$P:}
(a) Crystal structure of Ti$_{2}$Te$_{2}$P stacked in hexagonal layers, in which blue, brown and pink balls identify Te, Ti and P atoms, respectively. The dotted-rectangles enclose quintuple layers.    (b) 3D bulk Brillouin zone (BZ) of single crystals and its projected hexagonal surface where the high-symmetry points are labeled. 
(c) First-principles calculation of the bulk bands  with inclusion of the spin-orbit coupling.}
\end{figure}
In this work, we investigate the electronic band structures of the (001) surface of Ti$_{2}$Te$_{2}$P using high-resolution vacuum ultraviolet ARPES  and  first-principles calculations. Our measurements reveal the presence of  anisotropic  surface Dirac cones at the $\bar{M}$ point and nodal-line-like features along the $\bar{\Gamma}$-$\bar{M}$ direction. The experimental results are in excellent agreement with the first-principles calculations.  Parity calculations reveal the band-inversions at the $\Gamma$ and the F point indicating that the system is  topologically rich.  The anisotropic Dirac cones have the potential to be used in technological applications due to the different mobilities in a Dirac cone.
 \begin{figure*}[ht]
	\centering
	\includegraphics[width= 0.95\textwidth]{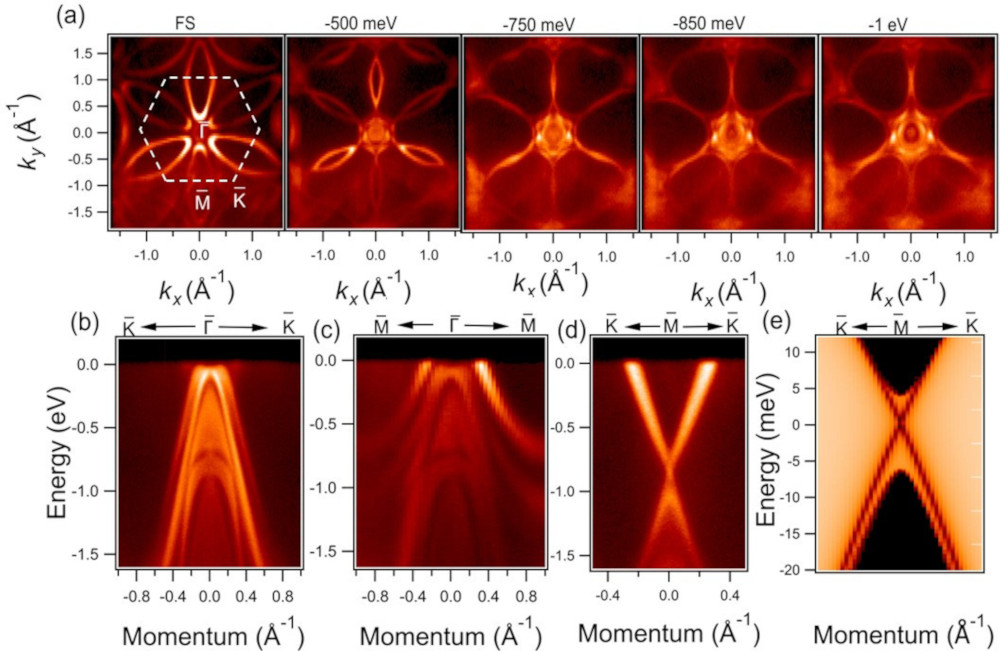}
	\caption{
	 {Fermi surface  map  and electronic band dispersions  of  Ti$_{2}$Te$_{2}$P}:
(a) Fermi surface map and constant energy contours at different binding energies measured at a photon energy of 100 eV. The white colored hexagonal shape is a constructed BZ in order to identify the high symmetry points.  Band dispersion along the (b) $\bar{K}$-$\bar{\Gamma}$-$\bar{K}$, (c) $\bar{M}$-$\bar{\Gamma}$-$\bar{M}$ and (d) $\bar{K}$-$\bar{M}$-$\bar{K}$ directions. {(e) Calculated band dispersion map along the  $\bar{K}$-$\bar{M}$-$\bar{K}$ direction. In the energy axis, the Dirac point is set as zero.} All the measured data were collected in  ALS beamline 4.0.3 at a temperature of 16 K.}
\end{figure*} \\
   Single crystals of Ti$_2$Te$_2$P were grown by chemical vapor transport method as described in Ref. \cite{Frauke, Oh}, and details of sample characterizations are described in the Supplemental Material (SM) note 1 \cite{SI}. 

Synchrotron-based ARPES measurements were performed at the Advanced Light Source (ALS) beamline 10.0.1 equipped with Scienta R4000, ALS beamline 4.0.3 equipped with R8000 hemispherical electron analyzers, and at the SIS-HRPES end-station at the Swiss Light Source (SLS) equipped with Scienta R4000. Similarly, helium lamp based ARPES measurements were performed at the
Laboratory for Advanced Spectroscopic Characterization of Quantum Materials (LASCQM) with R3000 hemispherical analyzer at University of Central Florida. The angular  and energy resolutions were  set to be better than 0.2$^{\circ}$ and 20 meV, respectively. The electronic structure calculations and structural optimization were carried out within the density-functional formalism, which are described in SM Note 2 \cite{SI,vasp, vasp1, pbe1, pbe2, pbe3}.
\begin{figure*}
	\centering
	\includegraphics[width=0.95\textwidth]{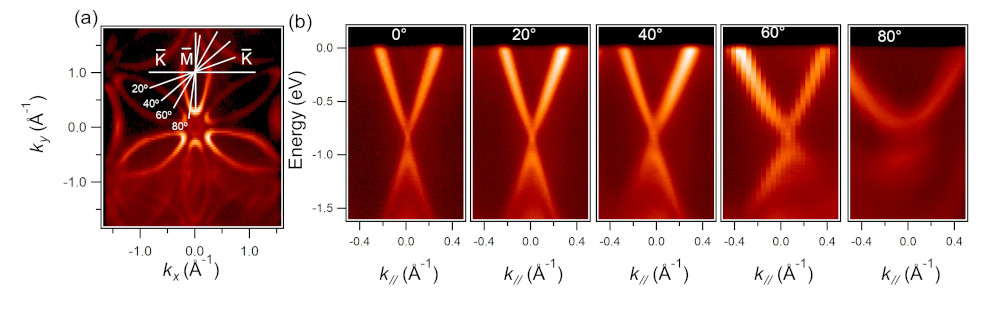}
	\caption{{Experimental observation of anisotropic Dirac cone}: (a) Fermi surface map in which white-colored lines are drawn via the $\bar{M}$ point to show  angles of dispersion maps with respect to the $\bar{K}$-$\bar{M}$-$\bar{K}$ direction.  
(b) Dispersion maps along different orientions, angles with respect to $\bar{K}$-$\bar{M}$-$\bar{K}$ direction are noted  on the top of plots.  All data
were collected at the beamline 4.0.3 in ALS at a photon energy of 100 eV.}
\end{figure*} 

We begin our discussion by presenting the crystal structure of Ti$_{2}$Te$_{2}$P in Fig. 1(a
). It is isostructural to  the  Bi$_{2}$Te$_{2}$Se, which crystallizes in a rhombohedral crystal structure
with  lattice parameter a = 3.6387(2) \AA , c = 28.486(2) \AA ~and space group $R\bar{3}m$ (No. 166), each unit cell consisting of three quintuple atomic layers separated by van der Waals gaps \cite{Frauke}. The  phosphorus and tellurium atoms are arranged in hexagonal layers stacked along the hexagonal c-axis in a sequence with two layers of tellurium followed by a single layer of phosphorus and the gap is filled by titanium as shown in Fig. 1(a). The (001) plane is the natural cleaving plane in this compound due to the atomic layers along the c-axis. A three dimensional BZ  and  its projection to 2D along the [001] direction are presented in Fig. 1(b), which illustrates the hexagonal symmetry of the (001) surface BZ. 
 Figure 1(c) shows the bulk band calculation of  Ti$_{2}$Te$_{2}$P (with spin-orbit coupling (SOC)) along the high symmetry directions which includes four time reversal invariant momenta (TRIM) points. 
 It shows a hole-like band around the $\Gamma$ point. A band is seen crossing the chemical potential at the $\Gamma$ point indicating the semimetallic behavior, which is flat in the vicinity of the Fermi level at the $\Gamma$ point. 
   Our detailed electronic structures calculations  show the occurrence of bulk  band inversion  both at the $\Gamma$ and the F point.  These band inversions are between  Ti-\textit{p} and Te-\textit{d}  states and involve different sets of bands (see SM \cite{SI}).  
 A bulk band inversion at the $F$ point takes place between
Ti \textit{p} and Te \textit{d} bands around 750 meV below the Fermi level,
which agrees well with both the calculated and observed
Dirac surface state at the $\bar{M}$ point. This surface state is of
the strong TI type, since our calculated $Z_2$  indices for this case are
(1;000) (see SM \cite{SI}).   Unlike the other sister compounds of this family (Zr$_{2}$Te$_{2}$P, Hf$_{2}$Te$_{2}$P), Ti$_{2}$Te$_{2}$P does not possess band inversion above the chemical potential at the zone center. The parity analysis of this system shows that a strong TI state exist at the $\Gamma$  point and at the $M$ point (see SM for the parity calculation \cite{SI}).  Our calculations indicate that the band inversion between different bulk bands at  the $\Gamma$ point also leads to strong TI topology with $\mathbb{Z}_2$ again (1;000). Remarkably, this band inversion takes place near a similar binding energy as the one responsible for the Dirac surface state at the $\bar{M}$ point (see \cite{SI}).\\
Figure 2 illustrates the detailed electronic structures of experimentally measured Ti$_{2}$Te$_{2}$P by using synchrotron based high resolution ARPES. In Fig. 2(a), the Fermi surface map is presented which has a six-fold flower petal shape as a consequence of the presence of the three-fold rotational and inversion symmetry. At the $\bar{\Gamma}$ point, a circular-shaped energy pocket can be seen due to the hole-like band that crosses the chemical potential. Furthermore, a constructed BZ is presented  in which high-symmetry points are  labeled.  {The photoemission spectra obtained from ARPES measurements suggests that} petal shaped Fermi pockets are  surface originated as they  do not disperse significantly with  photon energies (see SM \cite{SI}). The impacts of the matrix elements effects are noticeable leading to  non-uniform spectral intensities. Furthermore, petal-like features coming from the secondary BZ can be observed at the edges.  The ARPES measured Fermi surface of  Ti$_{2}$Te$_{2}$P  is similar to  Fermi-surface of the Zr$_{2}$Te$_{2}$P and Hf$_{2}$Te$_{2}$P \cite{Cava221, Niraj, Hosen221}.  The energy contours delineate how the band dispersions evolve with the binding energies. The energy pockets at the $\bar{\Gamma}$ point grow bigger with binding energy indicating the hole nature of the bands at the $\bar{\Gamma}$ point.  This hexagonal BZ has six $\bar{M}$-points at the center of the elliptical Fermi pockets. The diameters of the elliptical petal diminish with  binding energies. At the binding energy of 750 meV, the elliptical features collapse  into lines. Furthermore, the line-like feature seen along the $\bar{\Gamma}$-$\bar{M}$ direction form nodal-line-like states as it is the crossing point of bulk Dirac dispersions (see SM \cite{SI}). Interestingly, there are an even number of topological nontrivial surface states at the $\bar{\Gamma}$ point and one at the $\bar{M}$ point, which provide the condition for Dirac-node arcs \cite{Hosen221}. However, we do not observe  gapped bulk Dirac cones in the vicinity of the $\bar{\Gamma}$ as seen in  Hf$_2$Te$_2$P, probably due to weaker spin-orbit coupling, or due to energy resolution limitations obscuring the gap in the photoemission intensity plot. The node arc feature is protected by in-plane time-reversal invariance similar to its sister compound Hf$_2$Te$_2$P.  In order to study the dispersion maps along high symmetry axes, we present ARPES measurements of energy dispersion along high symmetry directions. Figure 2(b) shows  band structures along the $\bar{K}$-$\bar{\Gamma}$-$\bar{K}$ direction. Hole-like bands are seen in the vicinity of the Fermi level. Another band exists almost 750 meV below the Fermi level, which can be seen in the bulk band calculations. { Figure 2(c) displays the dispersion map along the $\bar{M}$-$\bar{\Gamma}$-$\bar{M}$, in which hole-like bands can be seen. The slab calculations (see SM \cite{SI}) suggest that most of the bands are bulk originated, whereas a pair of surface originated bands are buried within bulk bands.  } \\ 
  In Fig. 2(d), we present the dispersion map along the $\bar{K}$-$\bar{M}$-$\bar{K}$ direction in which, we observe a Dirac state at the binding energy of 750 meV. Our photon energy dependent measurements {suggest} the surface{-originated Dirac cone, { however the slab calculations (see Fig. 2(e)) suggests the presence of bulk bands. The bulk bands and surface bands, however, cannot be  resolved in the present resolution}.\\ 
 To show the anisotropic Dirac cones at the $\bar{M}$ points, we have taken dispersion maps along the different directions at the $\bar{M}$ point which have different Fermi velocities. Figure 3(a) shows the Fermi surface map in which   one $\bar{M}$ point is chosen as a reference point to demonstrate the anisotropy of the Dirac state at the $\bar{M}$ point. The Dirac cones are formed by the elliptical shaped petals at the Fermi surface. In Fig. 3(b), dispersion maps along the different directions are presented, which make angles of 0$^{\circ}$-80$^{\circ}$ at an interval of 20$^{\circ}$. The dispersion maps delineate the presence of anisotropic Dirac cones.\\
\begin{figure} 
\includegraphics[width= 0.95\textwidth]{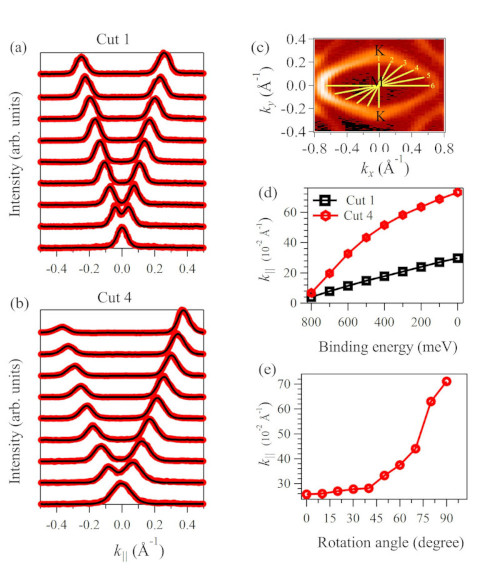}
\caption{{Demonstration of anisotropy in a Dirac cone}: (a)  Experimental momentum distribution curves (MDCs) [red open circles] integrated within 4 meV of the corresponding binding energy and fitted (black solid lines) to the experimental MDCs along cut 1 as shown in  (c). (b) Same as (a) but along cut 4 i.e, rotated by 60$^{\circ}$ with respect to the $\bar{K}$-$\bar{M}$-$\bar{K}$ direction. (c) Zoomed in view of the Fermi surface where the yellow lines 1-6 indicates cut directions in the interval of 20$^{\circ}$.  (d) The k$_{||}$ dispersions  as a function of binding energy. The co-ordinates are obtained from the MDCs plots from (a) and (b).  (e) The k$_{||}$ values as a function of rotation angles with respect to the $\bar{K}$-$\bar{M}$-$\bar{K}$ direction. }
\end{figure}
To  quantitatively determine the anisotropy of the Dirac cone in Ti$_2$Te$_2$P, we have performed a detailed analysis along both cut 1 i.e, along the $\bar{K}$-$\bar{M}$-$\bar{K}$ high-symmetry direction and cut 6 which is rotated 60$^{\circ}$ with respect to the $\bar{K}$-$\bar{M}$-$\bar{K}$ direction [see Figs. 4(a)-4(c)].  Figures 4(a)-4(b) show the momentum distribution curves (MDCs)(red open circles) which are extracted from the ARPES dispersions. In order to fit the obtained MDCs,  Gaussian-Lorentzian (GL) product functions have been used. The Gaussian part represents primarily the energy resolution of the instrument and the lifetime broadening of the photoelectrons,  whereas the core holes are captured by the Lorentzian width \cite{Valla1, Valla2}. The fitted MDCs as shown in Figs. 4a-b show accurate fits upto around 750 meV i.e, near the Dirac point. From the obtained results one can see that along cut 1, the dispersion of the topological surface states (TSSs) follows a linear behavior whereas along cut 4 the dispersion of the TSS deviated quite strongly from the linear behavior (see Fig. 4(d)). The different behavior of the dispersion of the TSS along the different directions thus clearly shows the anisotropic dispersion of the TSS. The magnitude of the Fermi velocity associated with cut 1 is almost double of cut 4. 
   Fig. 4(e) displays the change in the k$_{||}$ vector as a function of the rotation angle where the rotation has been performed with respect to the high symmetry direction $\bar{K}$-$\bar{M}$-$\bar{K}$; it clearly shows a significant increase of the k$_{||}$  wave vector above 40$^{\circ}$ rotation. \\
The  anisotropic surface Dirac cone in a 3D system has been overlooked  despite substantial works done on the tilted bulk Dirac cones in semimetallic systems.  
 The anisotropic Dirac cone at the $\bar{M}$ point of Ti$_2$Te$_2$P is the consequence of the petal shaped energy pocket at the $\bar{M}$ point in the BZ, due to which the Fermi velocities vary along the directions of the dispersion maps. 
  The anisotropy of  TSSs in some materials have been observed \cite{Kuroda, Chen, Souma, Rader, Zhang, Ru2Sn3}, which are  attributed to  the spin-dependent scattering giving rise to anisotropic scattering rates of the surface-state electrons, however the role of spin-dependent scattering  in Ti$_2$Te$_2$P is yet to be revealed. {Importantly, {materials with anisotropic surface Dirac cones could be a platform   that provide a direction dependent transport. It requires further explorations in  such materials. }   \\
 In conclusion, we observe  anisotropic Dirac cones at the $\bar{M}$ points in Ti$_2$Te$_2$P material. Our $\mathbb{Z}_2$ calculations identify this system as  topologically non-trivial. The presence of nodalline like features along the $\bar{\Gamma}$-$\bar{M}$ direction in addition to the anisotropic Dirac cones makes this system topologically rich. Our study provides an archetype system for the understanding of the 221-tetradymite system and we anticipate that it will stimulate further research interests in this class of materials.  \\
 
M.N. acknowledges  support from the Air Force Office of Scientific
Research under Award No. FA9550-17-1-0415, { the
Air Force Office of Scientific Research MURI (Grant
No. FA9550-20-1-0322)}, and
the National Science Foundation (NSF) CAREER award
DMR-1847962. A.K.N., A.A., and P.M.O. acknowledge support from the Swedish Research Council (VR) and from the Knut and Alice Wallenberg Foundation (Grant No. 2015.0060). Computational resources were provided by the Swedish National Infrastructure for computing (SNIC)(Grant No. 2018-05973). A. K. N. acknowledges the support of Department of Atomic Energy (DAE) and Science and Engineering Research Board (SERB) research grant (Grant No. SRG/2019/000867) of the Government of India.   We thank Sung-Kwan Mo and  Jonathan Denlinger for beamline assistance at the LBNL. This research used resources of the Advanced Light Source, a U.S. DOE Office of Science User Facility under contract no. DE-AC02-05CH11231.  We thank Nicholas
Clark Plumb for beamline assistance at the SLS, PSI.\\
$^*$ Corresponding author: Madhab.Neupane@ucf.edu

\newpage
\setcounter{figure}{0}
\begin{quote}
	\centering
\textbf{Supplementary Materials}
\end{quote}
\bigskip
\bigskip
\textbf{Supplementary Note 1: Synthesis and bulk characterization of   Ti$_2$Te$_2$P crystals}\\
\hspace{\parindent} Single crystals of Ti$_2$Te$_2$P were grown by the standard chemical vapor transport method.  The obtained crystals were thin platelets with typical dimensions 0.5 $\times$ 0.5 $\times$ 0.06 mm$^3$. They had dark metallic luster and were found moderately stable against air and moisture. Their chemical composition was checked by energy-dispersive X-ray analysis ( EDAX) using an FEI scanning electron microscope equipped with an EDAX Genesis XM4 spectrometer. Results confirmed that the material is homogeneous single-phase with the stoichiometry close to the nominal one (see Fig. S1). The crystal structure was verified by single crystal X-ray diffraction performed on a Kuma-Diffraction KM4 four-circle diffractometer equipped with a CCD camera using Mo K$\alpha$ radiation. 
The results confirmed the rhombohedral space group R$\bar{3}$m, and yielded the lattice parameters similar to those reported in the literature \cite{Frauke, Oh}. Electrical resistivity measurements were carried out within the temperature range of 2-300 K using a conventional four-point ac technique implemented in a Quantum Design PPMS platform. The electrical contacts to as-grown crystals were prepared using silver epoxy paste. The electrical resistivity was measured with electrical current flowing within the hexagonal plane of the Ti$_2$Te$_2$P crystal lattice.\\

Figure S2 shows the temperature dependence of the electrical resistivity of single-crystalline Ti$_2$Te$_2$P, measured within the hexagonal plane of the crystallographic unit cell. The compound exhibits metallic characteristics in regards to electrical transport, in concert with the previous report \cite{Oh}. However, at the lowest temperatures, the resistivity saturates at a constant value that can be associated with the residual scattering of conduction electrons on structural imperfections, at odds with the observation in Ref. \cite{Oh} of a logarithmic upturn below 7 K that was attributed to weak localization or electron-electron interaction. Here, it should be noted that the specimen investigated in Ref. \cite{Oh} was an ultra-thin flake exfoliated from a bulk crystal using the scotch type method, while in the present case it was a parallelepiped cut from a larger single-crystalline piece. The different form of the samples explains the discrepancy in the magnitude of the resistivity measured at room temperature (ca. 114 versus 30 $\mu \Omega$-cm, in the present study and Ref. \cite{Oh}, respectively) and at 2 K (ca. 24.4 versus  3.5 $\mu \Omega$-cm, respectively). Furthermore, it clarifies the absence in the resistivity data displayed in the inset to Fig. S2 of any clear features due to quantum corrections. \\

\noindent \textbf{Supplementary Note 2: First-principles calculations of  Ti$_2$Te$_2$P}\\
\hspace{\parindent} Structural optimization was carried out within the density-functional formalism as implemented in the Vienna ab initio simulation package (VASP) \cite{vasp, vasp1}. Exchange and correlation were treated within the generalized gradient approximation (GGA) using the parametrization of Perdew, Burke, and Ernzerhof (PBE) \cite{pbe1}. The projector augmented wave (PAW) method \cite{pbe2, pbe3} was employed for the wave functions and pseudo potentials to describe the interaction between the ion cores and valence electrons. The lattice constants and atomic geometries were fully optimized and obtained by minimization of the total energy of the bulk system. The surface of the 2D crystal was simulated as a slab calculation within the supercell approach with sufficiently thick vacuum layers. In addition to the general scalar-relativistic corrections in the Hamiltonian, the spin-orbit interaction was taken into account. The plane wave cutoff energy and the k-point sampling in the BZ integration were checked carefully to assure the numerical convergence of self-consistently determined quantities.\\

\noindent \textbf{Supplementary Note 3: Surface and bulk bands in Ti$_2$Te$_2$P}\\
\hspace{\parindent} Fermi surface maps at different photon energies confer an opportunity to study the surface nature of the Fermi pockets. Surface bands do not disperse with the photon energy. In  Fig.\ S3, it is evident that the petal-like energy pockets do not change their shape and size with change in  photon energies which indicates that the bands originate from the surface. Figure \ S4 illustrates the ARPES measured dispersion maps and their corresponding slab calculations  along the high symmetry directions $\bar{M}$-$\bar{\Gamma}$-$\bar{M}$ (Fig.\ S4(a)) and $\bar{K}$-$\bar{M}$-$\bar{K}$ (Fig.\ S4(b)), respectively.  In the vicinity of the $\bar{\Gamma}$ point as shown in Fig.\ S4(a), a pair of hole-like surface bands  is present along with bulk bands. One band crosses  the Fermi level, whereas, other  bands remain below the Fermi level. Ultimately, they merge at a certain $k$-point between the  $\bar{\Gamma}$ and the $\bar{M}$ point to become a single band.   Unlike other sister compounds of this family, it does not have a topological insulating state above the Fermi level. 
As discussed in the main text, the anisotropic Dirac cone is clearly seen on the theoretical plot shown in Fig.\ S4(b) right panel. The upper Dirac cone extends upto $\sim$750 meV above the Fermi level.
Photon energy dependent dispersion maps along the high symmetry directions are presented in Fig. \ S5. It suggests that a  surface state is seen at the $\bar{M}$ point (see Fig. S5(c)).
We present second derivatives, momentum distribution curves ( MDCs), Energy distribution curves ( EDCs) of dispersion maps along the high symmetry directions in Fig. \ S6 which provide the visualization of the bands in the respective directions.\\

\noindent \textbf{Supplementary Note 4: Anisotropic Dirac cones}\\
\hspace{\parindent} As seen in the main text, the Fermi velocity varies on the direction with respect to the $\bar{K}$-$\bar{M}$-$\bar{K}$ direction. Here, we show how Fermi velocities vary when they make angles of 60$^{\circ}$ and 30 $^{\circ}$ with respect to  the $\bar{K}$-$\bar{M}$-$\bar{K}$ direction (see Fig. S7). We take different $\bar{M}$ points and show how the dispersion maps evolve with the angles. The slopes of each dispersion map that makes 60$^{\circ}$-angles  with $\bar{K}$-$\bar{M}$-$\bar{K}$ should have the same magnitude, however, matrix elements effects change  the value  to some extent. Nevertheless, anisotropic behaviors are clearly seen.\\

\noindent \textbf{Supplementary Note 5: Observation of Dirac node-arc}\\
\hspace{\parindent}Observation of Dirac node-arc is presented in Fig. S8; we take 7 cuts in equal intervals perpendicular to the $\bar{\Gamma}$-$\bar{M}$ direction. Figure S8(b) shows the evidence of Dirac node-arc as all the dispersion maps show Dirac dispersions. The Dirac node-arc extends over more than 1 $\AA^{-1}$ in the $k$-space.\\

\noindent \textbf{Supplementary Note 6: Electronic structures of  Hf$_2$Te$_2$P and Zr$_2$Te$_2$P}\\
\hspace{\parindent}In order to investigate the electronic structures of sister compounds of Ti$_2$Te$_2$P, we present electronic structures of  Hf$_2$Te$_2$P and Zr$_2$Te$_2$P and compare them with  Ti$_2$Te$_2$P in Figs.\ S9-S11. ARPES spectra show  similar Fermi maps, however, they are topologically different as a consequence of spin-orbit coupling strength. Furthermore, Dirac points are located at different binding energies. In Fig. \ S9, the Fermi map and dispersion maps along high symmetry direction are presented. Even though many  features are similar to Ti$_2$Te$_2$P, some distinct features are seen in Hf$_2$Te$_2$P (for detail see Ref. \cite{Hosen}). Figure\ S10 displays ARPES measured electronic structures of Zr$_2$Te$_2$P, it also shows similar Fermi map however dispersion maps show somewhat different bands. Figures \ S11  show experimental dispersion maps along the high symmetry directions which illustrate the differences among these sister compounds.\\ 

\noindent \textbf{Supplementary Note 7: $\mathbb{Z}_{2}$ indices calculations}\\
\hspace{\parindent} In order to understand the topology of Ti$_2$Te$_2$P, we calculate  $\mathbb{Z}_{2}$ indices for different bands as shown in Fig. S12 and tabulate in the table I. The bands show different topological properties. Band 5 provides the topology of  the experimentally observed surface Dirac cone at the $\bar{M}$ (bulk F point) point which shows the evidence of strong TI. Similarly, Fig. S13 (and table II) displays the  $\mathbb{Z}_{2}$ indices of Hf$_{2}$Te$_{2}$P. Therefore, one can conclude Ti$_2$Te$_2$P is topologically different from Hf$_2$Te$_2$P. 
\newpage
\begin{figure*}
\centering
	\includegraphics[width=0.7\textwidth]{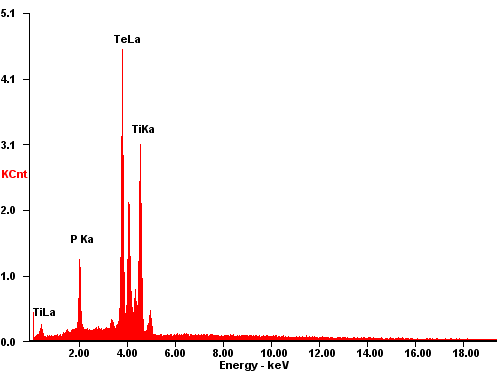}
	\begin{tabular}{|p{4cm}|p{4cm}|p{4cm}|} 
 \hline
Element &  Wt\% &  At\%\\
 \hline
 PK & 07.99 & 19.97  \\
 \hline
 TeL & 68.08 & 41.33\\
 \hline
 TiK & 23.93 & 38.70\\
 \hline
 Matrix &  correction & ZAF \\
 \hline
 \end{tabular}
	\caption{ Energy dispersive X-ray (EDX) spectroscopy analysis of Ti$_2$Te$_2$P.
	}
\end{figure*}
\begin{figure*}[h!]
	\centering
	\includegraphics[width=\textwidth]{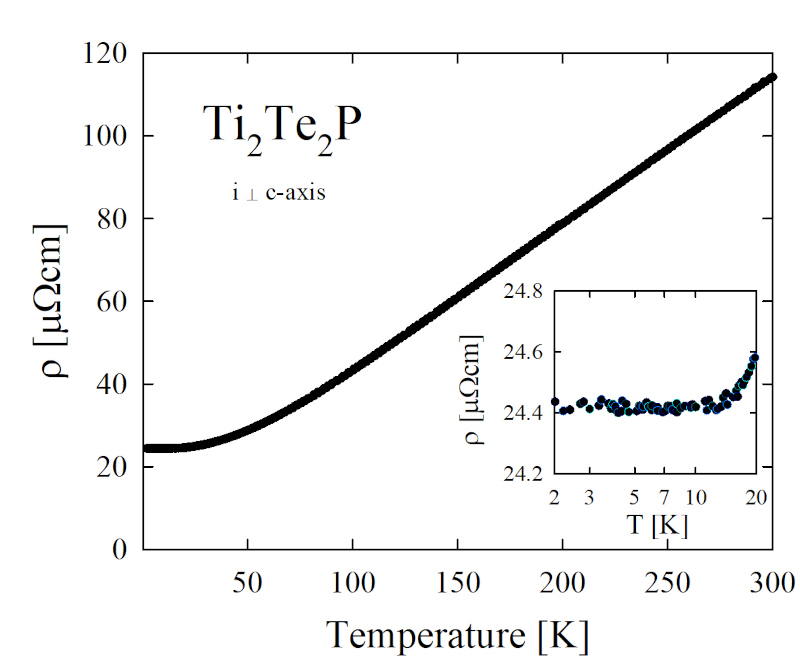}
	\caption{{Resistivity plot of Ti$_{2}$Te$_{2}$P.}
	Temperature variation of the electrical resistivity measured with electric current flowing within the hexagonal plane of the crystal. The inset shows the low temperature resistivity data.}
\end{figure*}
\begin{figure*}[h!]
	\centering
	\includegraphics[width=\textwidth]{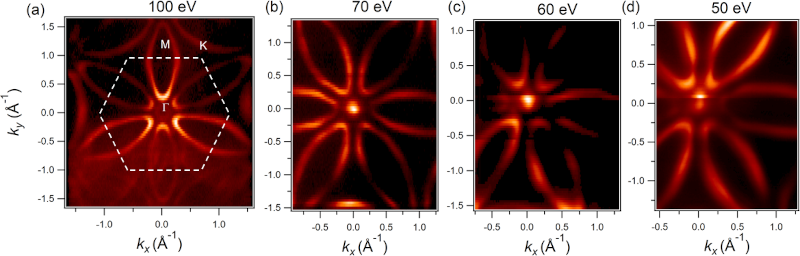}
	\caption{{Photon energy dependent Fermi surface.}
	(a)-(d) Photon energy dependent Fermi surface map.   The measurement shown in (a) was performed at the ALS beamline 4.0.3 at a temperature of 16 K  and (b)-(d) were performed at ALS beamline 10.0.1.}
\end{figure*}
\begin{figure*}[h!]
	\centering
	\includegraphics[width=\textwidth]{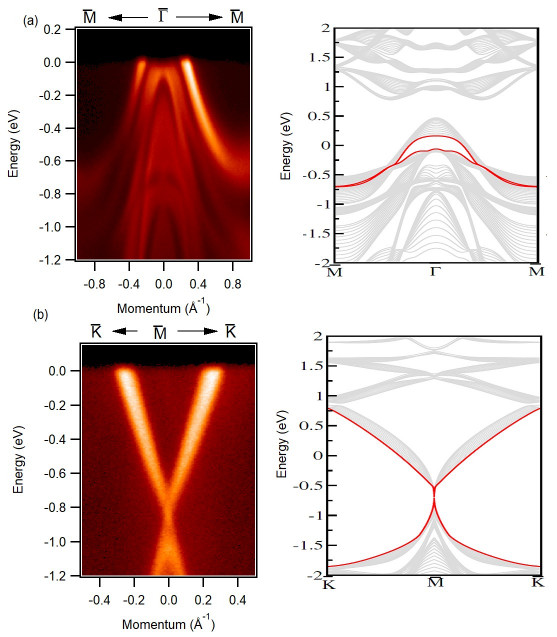}
	\caption{{Dispersion maps along the different high symmetry axes.}
(a)-(b) Band dispersions along the high symmetry directions as indicated on the top of the plots. Right panel plots are calculated slab calculations of corresponding left plots.}
\end{figure*}	 
\newpage
\noindent

\begin{figure*} [h!]
	\centering
	\includegraphics[width=\textwidth]{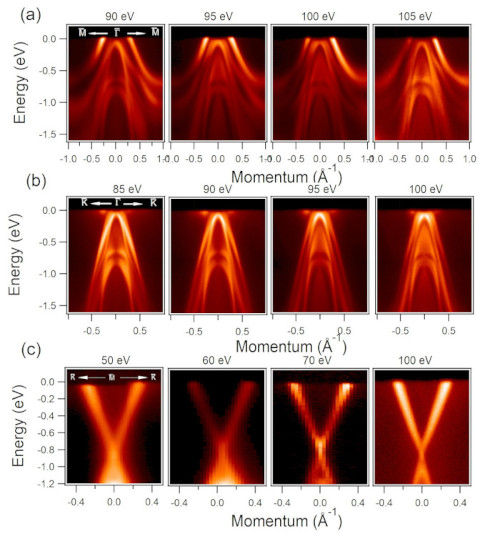} 
	\caption{{Photon energy dependent measurements along the high-symmetry directions.}
	 Photon energy dependent band dispersion along the (a) $\bar{M}$-$\bar{\Gamma}$-$\bar{M}$, (b) $\bar{K}$-$\bar{\Gamma}$-$\bar{K}$, and (d)  $\bar{K}$-$\bar{M}$-$\bar{K}$ directions, respectively. Photon energies  are labeled on the top of each plot. Measurements of dispersion maps shown in (a)-(b) and rightmost plot of panel (c)   were performed at ALS beamline 4.0.3 at a temperature of 18 K, and first three left plots of the panel c were performed  at the ALS beamline 10.0.1.}
\end{figure*}
\newpage
\noindent

\begin{figure*} [h!]
	\centering
	\includegraphics[width=\textwidth]{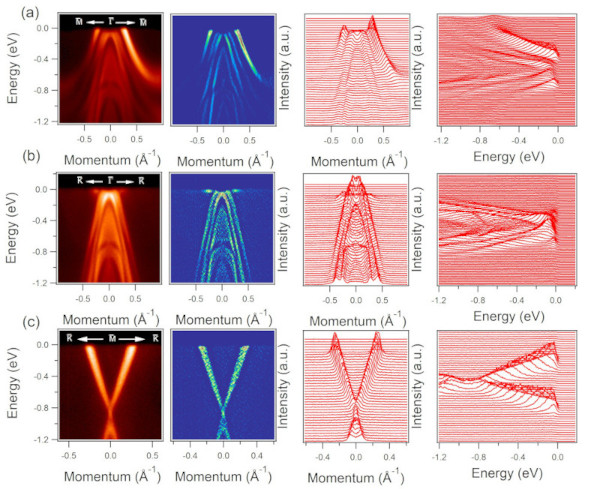} 
	\caption{{Dispersion maps along high-symmetry  directions, respective second derivative plots, MDCs and EDCs.}
	(a) Energy dispersion map, second derivative plot, MDC and EDC for the dispersion map along the (a) $\bar{M}$-$\bar{\Gamma}$-$\bar{M}$, (b) $\bar{K}$-$\bar{\Gamma}$-$\bar{K}$, and (c)  $\bar{K}$-$\bar{M}$-$\bar{K}$ direction, respectively.  Measurements were performed at ALS beamline 4.0.3 at a temperature of 16 K at a photon energy of 100 eV.
	}
\end{figure*}

\newpage
\noindent

\begin{figure*} [h!]
	\centering
	\includegraphics[width=\textwidth]{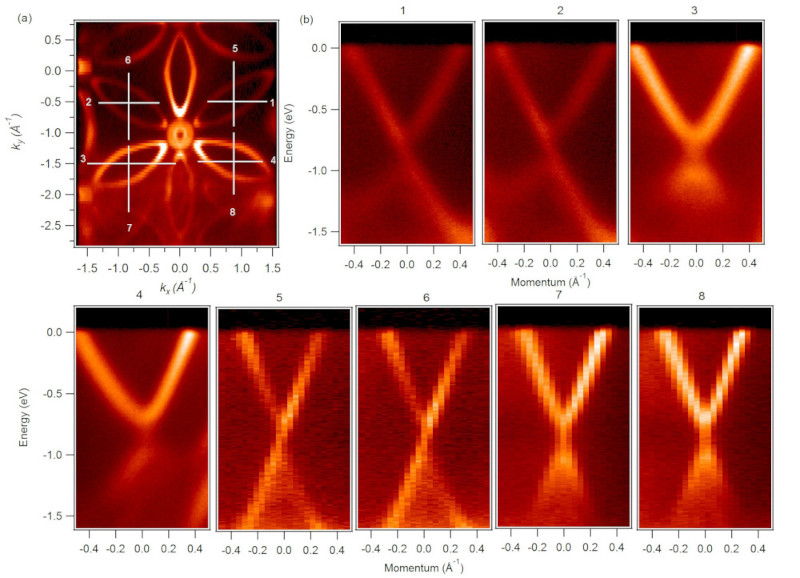} 
	\caption{{Observation of anisotropic Dirac cone at the $\bar{M}$ point.}
	(a) Fermi surface maps at a photon energy of 90 eV. Lines indicated by numbers represent the cut directions making angles of 60$^{\circ}$ and 30$^{\circ}$ alternately with the $\bar{K}$-$\bar{M}$-$\bar{K}$ directions at different M points in  the Brillouin zone.
 (b) Dispersion maps  for the cuts indicated on (a). Measurements were performed in ALS beamline 4.0.3 at a temperature of 16 K at a photon energy of 100 eV.
   	}
\end{figure*}

\begin{figure*} [h!]
	\centering
	\includegraphics[width=\textwidth]{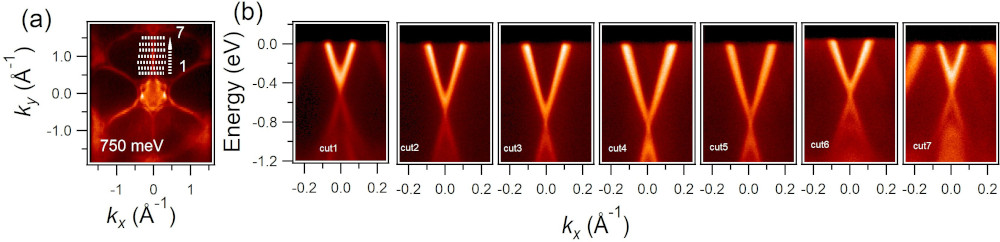} 
	\caption{{Observation of node arc along the $\bar{\Gamma}$-$\bar{M}$ direction.}
	(a) Constant energy contour at a binding energy of 750 meV. Dotted lines show the cut directions. (b) Cuts are taken perpendicular  to $\bar{\Gamma}$-$\bar{M}$-$\bar{\Gamma}$ direction as shown in (a). Measurements were performed in ALS beamline 4.0.3 at a temperature of 16 K at a photon energy of 100 eV.
 	}
\end{figure*}
\newpage
\begin{figure*} [h!]
	\centering
	\includegraphics[width=\textwidth]{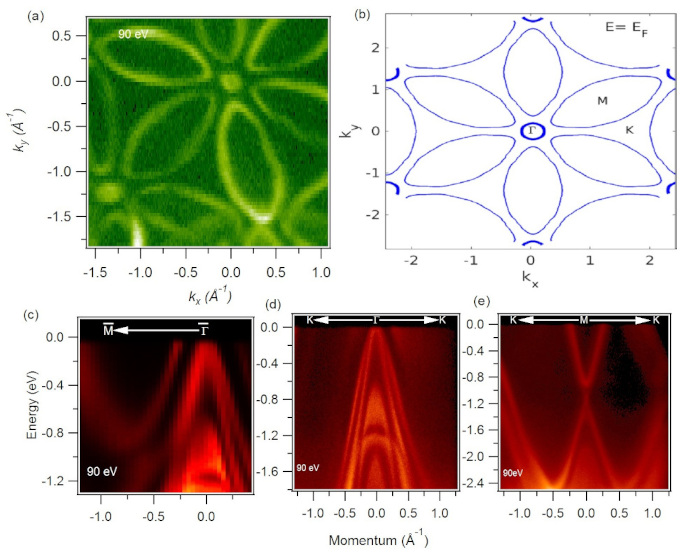} 
	\caption{{Fermi surface and dispersion maps along the high-symmetry directions in Hf$_2$Te$_2$P.}
	(a) Fermi surface maps  measured at a photon energy of 90 eV.
 (b) Calculated Fermi surface.
    Dispersion maps measured along the (c) $\bar{M}$-$\bar{\Gamma}$-$\bar{M}$, (d)   $\bar{K}$-$\bar{\Gamma}$-$\bar{K}$, (e)   $\bar{K}$-$\bar{M}$-$\bar{K}$ directions, respectively.  Data were
collected at the SIS-HRPES end station at the SLS, PSI at a temperature of 18 K.}
\end{figure*}
\newpage
\begin{figure*} [h!]
	\centering
	\includegraphics[width=\textwidth]{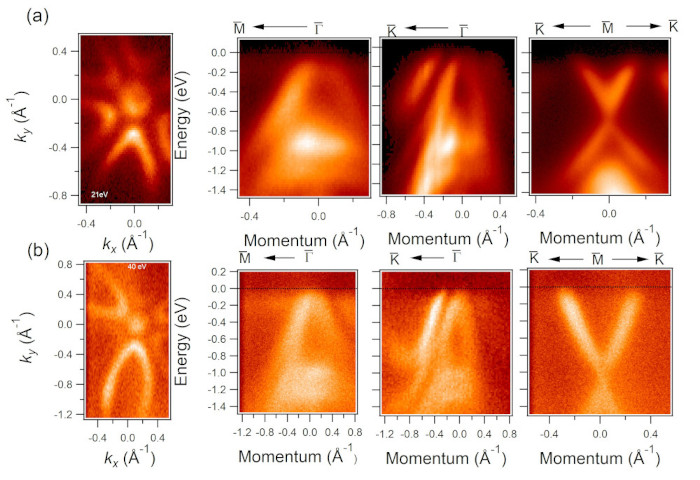} 
	\caption{{Fermi surface and dispersion maps along the high-symmetry directions in Zr$_2$Te$_{2}$P.}
	 Fermi surface map and  dispersion maps measured along the $\bar{M}$-$\bar{\Gamma}$, $\bar{K}$-$\bar{\Gamma}$ and $\bar{K}$-$\bar{M}$-$\bar{K}$ direction at a photon energy of (a) 21 eV and (b) 40 eV.  Data were
collected at the Laboratory for Advanced Spectroscopic Characterization of Quantum Materials (LASCQM), UCF at a temperature of 20 K.	}
\end{figure*}

\begin{figure*} [h!]
\centering
\includegraphics[width= \textwidth]{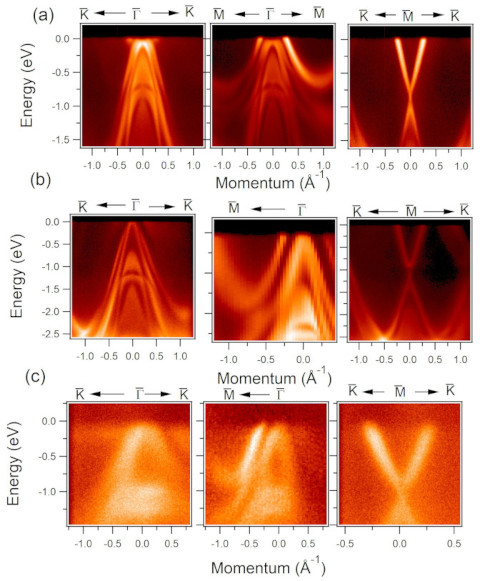}
\caption{{Measured electronic band structures of  Ti$_{2}$Te$_{2}$P, Hf$_{2}$Te$_{2}$P and Zr$_{2}$Te$_{2}$P.}
 Dispersion maps along the high-symmetry directions are shown in (a) for Ti$_{2}$Te$_{2}$P, (b) Hf$_{2}$Te$_{2}$P and (c) Zr$_{2}$Te$_{2}$P, respectively. }
\end{figure*}

\clearpage

\noindent

\begin{figure*}[ht] 
\centering
\includegraphics[width=\textwidth]{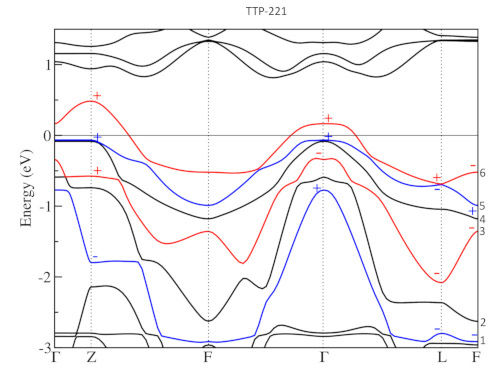} 
\caption{ {Parities of bands at TRIM points in Ti$_{2}$Te$_{2}$P.}  Even (odd) parity is labeled + (-). Six bands are taken for the $\mathbb{Z}_{2}$ calculations.}
\end{figure*}

\begin{table*} [h!]
\centering
\begin{tabular}{|p{2cm}|p{3cm}|p{2cm}|p{2cm}| } 
 \hline
  & & & \\
 Band & Z$_{2}$ invariants & $\widetilde{\Gamma}$ & $\widetilde{M}$ \\
 \hline
 1 & (1;000) & \checkmark  &   \\
 \hline
  2 & (1;000) & \checkmark  &   \\
 \hline 3 & (0;111) &   &   \\
 \hline 4 & (0;111) &   &   \\
 \hline 5 & (1;000) &   & \checkmark \\
 \hline 6 & trivial &   &   \\
 \hline
 \end{tabular}
 \caption{Calculations of $\mathbb{Z}_{2}$ indices in  Ti$_{2}$Te$_{2}$P.} 
 \end{table*}

\newpage
\begin{figure*} [h!]
\centering
\includegraphics[width=\textwidth]{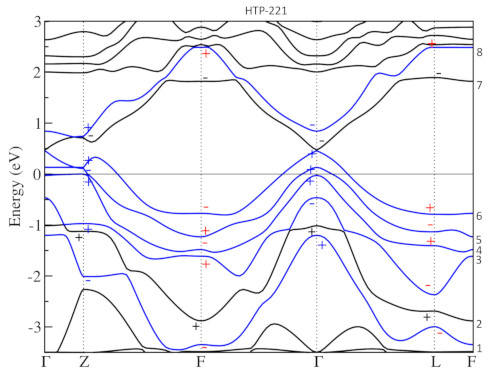}
\caption{{ Parities of bands at TRIM points in Hf$_{2}$Te$_{2}$P.}  Even (odd) parity is labeled + (-). Eight bands are taken for calculations.}
\end{figure*}

\begin{table*} [h!]
\centering
\begin{tabular}{|p{2cm}|p{3cm}|p{2cm}|p{2cm}| } 
 \hline
  & & & \\
 Band & Z$_{2}$ invariants & $\widetilde{\Gamma}$ & $\widetilde{M}$ \\
 \hline
 1 & (1;000) & \checkmark  &   \\
 \hline 
 2 & Trivial &   &   \\
 \hline 3 & (1;111) & \checkmark   & \checkmark  \\
 \hline 4 & (0;111) &   & \checkmark \\
 \hline 5 & (0;111)  & \checkmark  & \checkmark  \\
 \hline 6 & (1;111) &   & \checkmark \\
 \hline 7 & Trivial &   &  \\
 \hline 8 & (0;111) &   & \checkmark \\
 \hline
 \end{tabular}
 \caption{Calculations of $\mathbb{Z}_{2}$ indices in  Hf$_{2}$Te$_{2}$P.} 
 \end{table*}

\end{document}